# TITLE PAGE

# Reasoning Language Models for complex assessments tasks: Evaluating parental cooperation from child protection case reports

Dragan Stoll[a,b] (corresponding author), Brian E. Perron[c], Zia Qi[c], Selina Steinmann[b], Nicole F. Eicher [b], Andreas Jud[b, d]

[a] Institute of Psychology and Education, Ulm University, Germany

[b] School of Social Work, ZHAW Zurich University of Applied Sciences, Switzerland

[c] School of Social Work, University of Michigan

[d] Child and Adolescent Psychiatry, Psychosomatics, and Psychotherapy, Ulm University Clinics, Ulm, Germany

**Author Note**

**Dragan Stoll**, dragan.stoll@zhaw.ch, https://orcid.org/0009-0008-9088-5282, ZHAW School of Social Work, Pfingstweidstrasse 96, P.O. Box 707, CH-8037 Zurich, Switzerland, +41797645035

**Brian E. Perron**, beperron@umich.edu, https://orcid.org/0009-0008-4865-451X

**Zia Qi**, qizixuan@umich.edu, https://orcid.org/0000-0002-8407-0465

**Selina Steinmann**, selina.steinmann@zhaw.ch, https://orcid.org/0009-0004-5172-5009

**Nicole Florence Eicher**, nicoleflorence.eicher@zhaw.ch, https://orcid.org/0009-0006-3679-5986

**Andreas Jud**, andreas.jud2@zhaw.ch, https://orcid.org/0000-0003-0135-4196

**Competing interests**: The authors declare that they have no competing interests.

**Funding**: Our research was primarily funded by the Zurich Higher Education Institutions (DIZH) which aims to advance research and innovation on digitalization by using interdisciplinary approaches.

**Declarations Availability of data and materials**: The datasets generated and analyzed are not publicly available due to them containing information that could compromise the privacy of the research participants. However, programming code developed for this research is publicly available in the project GitHub repository (https://github.com/dragstoll/prompting_rag_methods_cm/tree/main/parental_cooperation_classification_reasoning)

**CRediT authorship contribution statement**

**Dragan Stoll:** Funding Acquisition, Project Administration, Conceptualization, Methodology, Software, Data Curation, Validation, Formal Analysis, Investigation, Visualization, Writing – original draft, Reviewing & Editing. **Andreas Jud:** Supervision, Conceptualization, Methodology, Writing – Reviewing & Editing. **Bryan E. Perron:** Conceptualization, Methodology, Writing – Reviewing & Editing. **Zia Qi:** Conceptualization, Methodology, Writing – Reviewing & Editing. **Selina Steinmann:** Conceptualization, Methodology, Validation. **Nicole Florence Eicher:** Conceptualization, Methodology, Validation.

# Reasoning Language Models for complex assessment tasks: Evaluating parental cooperation from child protection case reports

**Abstract**

**Purpose:** Reasoning language models (RLMs) have demonstrated significant advances in solving complex reasoning tasks. We examined their potential to assess parental cooperation during CPS interventions using case reports, a case factor characterized by ambiguous and conflicting information.

**Methods:** A four-stage workflow comprising (1) case reports collection, (2) reasoning-based assessment of parental cooperation, (3) automated category extraction, and (4) case labeling was developed. The performance of RLMs with different parameter sizes (255B, 32B, 4B) was compared against human-validated data. Two expert human reviewers (EHRs) independently classified a weighted random sample of reports.

**Results:** The largest RLM achieved the highest accuracy (89%), outperforming the initial approach (80%). Classification accuracy was higher for mothers (93%) than for fathers (85%), and EHRs exhibited similar differences.

**Conclusions:** RLMs' reasoning can effectively assess complex case factors such as parental cooperation. Lower accuracy in assessing fathers' cooperation supports the argument of a stronger professional focus on mothers in CPS interventions.

**Reasoning Language Models for complex assessment tasks: Evaluating parental cooperation from child protection case reports**

Parental cooperation during child protection services (CPS) interventions directly influences case trajectories, intervention success, and child safety outcomes (Ben-David, 2016; Charest-Belzile et al., 2020). Cooperation encompasses the mutual, intentional, and behavioral involvement of caregivers in services provided by child protection and related agencies (Platt, 2012). Consequently, its absence limits parents' capacity to make necessary changes, reducing the likelihood of positive outcomes in CPS interventions (Ben-David, 2016). The practical implications extend beyond mere compliance—"doing what they are told"—toward "true engagement" (Charest-Belzile et al., 2020; Platt, 2012); beyond acting out of fear of legal consequences (Ben-David, 2016); and beyond disguised compliance, concealing underlying parental opposition (Forrester et al., 2012; Mason et al., 2020). Effective cooperation requires *attitudinal* engagement, demonstrated through trust and openness toward services; *relational* alignment, characterized by a collaborative working relationship with caseworkers; and *behavioral* participation, expressed through consistent attendance and active involvement in interventions (Charest-Belzile et al., 2020). Moreover, drawing on our research experience, we consider parental cooperation to be an important, though often implicit goal, alongside other core objectives, in CPS interventions.

Given its recognized importance, parental cooperation has increasingly come to the forefront in social work practice and has received substantial research attention in both qualitative and quantitative studies (Gautschi, 2021; Jud & Gartenhauser, 2015; Lätsch et al., 2022). The proposed automated classification method would enable population-level data collection on this case factor. Unlike discrete variables such as substance use or domestic violence, cooperation is not captured in the structured administrative data maintained by child welfare information systems. Information about cooperation exists only in narrative documentation: case notes, assessment reports, and accountability reports written by caseworkers over the course of an intervention (Witte, 2020). This presents a fundamental measurement challenge. Cooperation is inherently dynamic; it emerges, fluctuates, and evolves across months or years of service involvement. A parent who initially resists engagement may develop genuine collaboration over time, while apparent compliance may mask underlying resistance (Forrester et al., 2012; Mason et al., 2020). Capturing these patterns requires analyzing narrative records that document changing circumstances over extended intervention periods. Systematic assessment of cooperation from case documentation presents distinctive analytical challenges. Information within a single case report is often contradictory, ambiguous, or reflects change over time. Assessment, therefore, requires weighing conflicting evidence and determining the net trajectory rather than making simple present/absent classifications.

**Gender perspective of parental cooperation**

Additionally, research on gender-related perspectives on working with parents in child welfare suggests that professionals interact differently with mothers and fathers, leading to a systematic bias in cooperation assessments (Philip et al., 2019; J. Scourfield et al., 2024; J. B. Scourfield, 2001). Child protection systems are characterized as mother-centric (Strega et al., 2008), assigning mothers primary responsibility for children's well-being while often treating fathers as peripheral, non-clients (J. Scourfield et al., 2024). Social workers tend to be more tolerant of non-cooperation from sole-carer mothers, who are seen as having no choice but to stay engaged for the child's benefit (Philip et al., 2019). Mothers are contacted earlier and more often, while information on fathers is frequently indirect due to rushed or delayed engagement (Dominelli et al., 2011; Philip et al., 2019; Strega et al., 2008). Organizational routines reinforce this imbalance, resulting in case documentation that provides more detailed and nuanced information about mothers than fathers (Strega et al., 2008), while fathers are often assessed in binary terms—cooperative or not—rather than through nuanced evaluation (Philip et al., 2019).

**Automated text analysis in child welfare through reasoning methods**

Recent advances in natural language processing have expanded the capacity to extract structured data from unstructured case narratives, driven in part by the need for secure methods that protect highly sensitive child-welfare data from disclosure or use by commercial AI providers. Language models (computational systems trained to process and generate human language) can now classify discrete risk factors in child welfare records with accuracy comparable to human experts (Perron et al., 2024; Qi et al., 2026). They are available as open-source software and can be deployed locally within the secure environment of a child protection agency or research institution, ensuring that all data remain on-site and comply with data-protection requirements. These methods have successfully identified parental substance-related problems, domestic violence, opioid involvement, and firearms presence without requiring extensive manual annotation of training data. However, standard approaches struggle when information is divergent, shifting, or requires synthesis across extended narratives. Our prior attempt to classify parental cooperation using retrieval-augmented generation (RAG), a technique that retrieves relevant text segments before classification, achieved accuracy of 80%, which could be further improved (Stoll et al., 2025). Error analysis revealed that the method failed to reconcile contradictory evidence within reports or to track how cooperation evolved across the reporting period.

Reasoning language models (RLMs) offer a methodological advance suited to this analytical challenge. These models are designed to decompose complex problems into sequential steps before producing final outputs (Besta et al., 2025; Raschka, 2025). The core mechanism, termed chain-of-thought processing, encourages the model to generate intermediate reasoning steps that lead to a conclusion rather than producing direct classifications (Wei et al., 2022). For cooperation assessment, this means the model can identify evidence both for and against cooperation, weigh temporal changes documented in the narrative, and reconcile contradictions before rendering a final determination. This process mirrors the cognitive approach that expert human reviewers employ when evaluating complex case information. Recent research demonstrates that reasoning-enabled models achieve substantial performance improvements on classification tasks in child welfare contexts, with smaller models matching or exceeding the accuracy of much larger architectures when reasoning capabilities are enabled (Qi et al., 2026).

**Purpose**

This study evaluates whether reasoning language models can accurately assess parental cooperation from CPS case reports, a construct marked by ambiguous and conflicting information that has proven difficult for prior automated approaches. We pursue three objectives. First, we compare reasoning-enabled models of different sizes (4 billion to 255 billion parameters) against our prior RAG approach to determine whether reasoning capabilities improve classification accuracy for this complex task. Second, using a semantic approach designed to distinguish caregiver roles, we assess whether model performance differs between mothers and fathers, informed by hypotheses from research on gendered documentation patterns in child welfare. Third, we demonstrate practical utility by applying the validated method to classify cooperation across a large corpus of CPS cases (N = 29,770 reports spanning 12,607 cases), generating structured data that would be infeasible to obtain through manual review. The findings address both methodological questions about appropriate AI techniques for complex assessment tasks and substantive questions about how cooperation is documented and understood across caregiver roles in child protection practice.

## Materials and methods

**Data source**

This study utilizes casework reports from the child protection system in the Canton of Zurich, Switzerland. In this system, Child Protection Authorities (comparable to child welfare courts in other countries) assess potential maltreatment cases and determine necessary interventions. The most common intervention involves appointing a social worker as an "assistant to the child," who provides counsel to families, coordinates services, and monitors case progress (Jud et al., 2011). Approximately every two years, appointed social workers submit accountability reports to the authorities documenting the case status, family circumstances, and intervention outcomes.

Since 2008, these reports have followed a standardized structure including: contact and demographic information; the legal mandate and case objectives; the child's current situation and well-being; case developments over the reporting period; perspectives of the child and parents on key matters; prognosis for future development; and recommendations for continued action. We obtained all 29,770 casework reports from 12,607 CPS cases spanning 2008 to 2022 through a data-sharing agreement with the cantonal authorities. Reports averaged approximately 1,300 words (five pages), with a standard deviation of 730 words. The median duration of CPS assistance was 3.3 years, though cases in the upper quintile averaged approximately ten years. Table 1 summarizes data source characteristics.

[INSERT TABLE 1 ABOUT HERE]

**Analytical workflow**

We developed a four-stage workflow for assessing parental cooperation using reasoning language models. The first stage involved collecting and preprocessing all casework reports from their original word processing format into plain text. The second stage applied reasoning-based question answering to assess maternal and paternal cooperation separately. The third stage extracted structured cooperation categories from model outputs. The fourth stage aggregated report-level classifications to generate case-level labels indicating whether lack of cooperation occurred at any point during CPS involvement.

**Reasoning-based assessment**

Reasoning language models differ from standard language models in how they process complex problems. Standard models generate outputs directly from learned patterns through single-pass processing. Reasoning models, by contrast, decompose problems into sequential steps before producing final answers, a technique termed chain-of-thought processing (Wei et al., 2022). Compared with instruction language models, they are given a larger "thinking budget" in the form of longer reasoning traces, which allows them to work through complex problems more thoroughly. For cooperation assessment, this means the model generates intermediate reasoning that identifies evidence for and against cooperation, weighs how cooperation evolved across the reporting period, and reconciles contradictory information before rendering a final decision.

We developed the assessment prompts through iterative refinement by examining model outputs on randomly sampled reports (Perron et al., 2024). The full prompt used in the study is provided in the supplementary material (Appendix A). It consists of five components: (1) an instruction specifying that the model should answer based only on explicitly documented information; (2) the assessment question with three response categories (lack of cooperation, cooperation present or emerged, no evidence); (3) operational definitions of lack of cooperation and established cooperation; (4) assessment guidelines addressing common interpretation challenges; and (5) the case report text. We assessed mothers and fathers cooperation separately using parallel prompts.

Lack of cooperation was operationally defined as: the parent does not follow professional instructions; is uncooperative or unwilling to work with the caseworker or other professionals; does not respond to or follow guidance; does not attend agreed appointments; cannot be motivated to adopt new perspectives; or is formally instructed under Article 307 of the Swiss Civil Code, which imposes a legal obligation to comply with child protection authority orders (Swiss Civil Code (Zivilgesetzbuch, ZGB), 1907). Whereas cooperation, present or emerging, was defined as: the parent is cooperative and willing to work with professionals, or willingness has developed over time; the parent responds to and follows professional guidance; and the parent attends agreed-upon appointments.

The assessment guidelines addressed interpretive challenges identified during prompt development. When reports contained evidence of both cooperation and lack of cooperation, the model was instructed to evaluate the overall trajectory to determine whether cooperation emerged over time. When no evidence of either cooperation or lack of cooperation was observed, the model was instructed to classify as "no evidence" rather than infer cooperation status. When reports referred to "parents" collectively, the model was instructed to apply this information to both the mother and father assessments.

**Model selection and configuration**

We evaluated open-source reasoning language models of three sizes to assess the relationship between model capacity and classification accuracy. Model capacity is measured in parameters (the numerical values learned during training that determine model behavior), with larger parameter counts generally associated with greater capability but also greater computational requirements. We selected models from the Qwen3 family (Yang et al., 2025) because these models demonstrated strong performance on multilingual tasks, including German text processing.

The three reasoning models evaluated were Qwen3-235B-A22B (255 billion total parameters, 22 billion active per inference), Qwen3-32B (32 billion parameters), and Qwen3-4B (4 billion parameters). The largest model employs a mixture-of-experts architecture, meaning it activates only specialized subnetworks, chosen experts, for each task rather than processing through all parameters simultaneously. For comparison with our prior approach, we also evaluated a 123-billion-parameter instruction-tuned model (Mistral-Large-Instruct-2411) using retrieval-augmented generation, which retrieves relevant text segments before classification, rather than processing complete reports through reasoning steps (Jiang et al., 2023).

All models were deployed using 4-bit quantization, a compression technique that reduces memory requirements while preserving most of the model's capabilities (Dettmers et al., 2023) and were obtained from the Hugging Face model repository. Following recommendations from the Qwen3 development team, we configured reasoning models with a temperature of 0.6, top-k of 20, and top-p of 0.95. Temperature controls output randomness, with lower values producing more deterministic responses; top-k and top-p control the diversity of word choices during generation. Maximum output length was set to 8,000 tokens (approximately 6,000 words; equivalent of up to 20 pages of text) to accommodate extended reasoning chains. Model inference was performed using the vLLM framework, which optimizes processing speed for large language models (Kwon et al., 2023).

**Category extraction and case labeling**

Reasoning model outputs contain two components: a "thinking" section that documents intermediate reasoning steps, and a final answer that includes the classification with supporting justification. An anonymized final assessment illustrating a case of non-cooperation is provided in Appendix B of the supplementary material. We extracted cooperation categories from the final answer using a separate instruction-tuned model (Qwen3-32B), configured to output structured JSON format. This extraction model received only the final answer text and was instructed to identify which of the three categories (lack of cooperation, cooperation present, or no evidence) the reasoning model selected. The extraction prompt is provided in Appendix C of the supplementary material.

For analysis purposes, we aggregated cooperation categories into a binary classification. During validation, expert human reviewers evidenced a difficulty distinguishing between two scenarios: cases in which documentation provided sufficient evidence to conclude that cooperation was present, and cases in which the documentation simply lacked any information about cooperation. Both scenarios share a common practical characteristic; neither contains evidence of problematic engagement. We therefore combined the "cooperation present or emerged" and "no evidence" categories into a single category indicating no documented lack of cooperation, while retaining "lack of cooperation" as a distinct category.

During the development and validation of our approach, we observed in the casework reports the binary framing that typically characterizes practitioners' assessments of parental cooperation; resistance or non-engagement triggers concern and is documented, while cooperative behavior is only sporadically mentioned, mostly in combination with other arguments regarding the case development. We conducted sensitivity analyses comparing model performance under the original three-category scheme versus the binary classification to assess whether this aggregation affected accuracy estimates. Case-level labels were then generated by flagging any case in which at least one report indicated a lack of cooperation by the respective parent.

**Validation procedures**

We validated model classifications against expert human review using a stratified random sample of 100 casework reports. The sample was constructed to ensure representation across classification combinations: 20 reports with both parents classified as lacking cooperation, 20 reports with neither parent classified as lacking cooperation, and 15 reports each for the two discordant patterns (mother lacking cooperation but not father; father lacking cooperation but not mother). Two expert human reviewers (EHRs), both with graduate training in social work and experience in child protection practice and documentation, independently classified maternal and paternal cooperation for each sampled report.

Expert reviewers were instructed to examine each report for indications of cooperation or lack thereof, note specific passages informing their evaluation, and classify cooperation for each parent using the same three-category scheme applied by the models. Reviewers provided justification for classifications when deemed necessary. All disagreements between reviewers were discussed and resolved through consensus, producing a benchmark dataset against which model performance was evaluated.

**Evaluation metrics**

Model performance was assessed using four metrics. Weighted overall accuracy measured the proportion of classifications matching the expert consensus benchmark. Precision measured the proportion of positive predictions (lack of cooperation) that were correct. Recall (also termed sensitivity) measures the proportion of actual positive cases that the model correctly identified. F1-score provided the harmonic mean of precision and recall, balancing both error types in a single measure.

We additionally computed Cohen's kappa (κ) to measure agreement after accounting for the level of agreement expected by chance. Following conventional interpretation guidelines (Landis & Koch, 1977), kappa values between 0.41 and 0.60 indicate moderate agreement, between 0.61 and 0.80 indicate substantial agreement, and above 0.81 indicate almost perfect agreement. We computed kappa for model-to-benchmark agreement and for inter-rater agreement between the two expert reviewers to contextualize model performance relative to human consistency. Confusion matrices were constructed to characterize the nature and direction of classification errors.

**Data protection**

Given the sensitive nature of child protection records, we implemented comprehensive data protection measures. All researchers signed confidentiality agreements. Structured data were anonymized by excluding or recoding identifying variables. All model processing was conducted locally on secure infrastructure without transmission to external servers. A detailed data management and protection plan was reviewed by the cantonal data privacy office and approved by a data privacy and ethics lawyer. The computational infrastructure consisted of high-performance computing server-class hardware with sufficient graphical memory to deploy the evaluated models locally. Due to confidentiality requirements, the datasets cannot be made publicly available; however, all analysis code is available in the project's GitHub repository.

## Results

**Reasoning models versus RAG approach**

Table 2 presents accuracy, precision, recall, and F1-score for classifying parental cooperation, comparing reasoning models of three sizes against the retrieval-augmented generation (RAG) approach. The large reasoning model (255B parameters) achieved the highest overall accuracy (89%), representing a nine percentage point improvement over the RAG-based approach (80%). Performance scaled with model size: the medium reasoning model (32B) achieved 84% accuracy, and the small reasoning model (4B) achieved 80% accuracy. The small reasoning model matched the performance of the substantially larger RAG-based instruct model (123B), suggesting that reasoning capabilities can compensate for reduced model capacity.

[INSERT TABLE 2 ABOUT HERE]

Table 3 presents confusion matrices for the large reasoning model, which achieved the highest overall performance and was subsequently used for full corpus classification. Across both parents (N = 200 classifications), the model correctly identified 118 of 132 actual cases of cooperation (89.4% recall) and correctly classified 60 of 68 cases without documented lack of cooperation (88.2% specificity). The false positive rate (11.8%) was lower than that of the RAG-based approach (25.0%), indicating that reasoning capabilities reduced the erroneous identification of cooperation problems where none existed.

[INSERT TABLE 3 ABOUT HERE]

Table 4 presents Cohen's kappa coefficients measuring agreement between model classifications and the expert consensus benchmark. The large reasoning model achieved κ = 0.76 for overall parental cooperation, indicating substantial agreement according to conventional interpretation guidelines (Landis & Koch, 1977). This represents an improvement over the prior RAG-based approach (κ = 0.62). To contextualize model performance relative to human consistency, we also computed kappa between the two expert human reviewers (EHRs) before consensus was reached. The model's agreement with the benchmark (κ = 0.76) approached the level observed between individual EHRs and the consensus benchmark (κ = 0.88 and κ = 0.80, respectively).

[INSERT TABLE 4 ABOUT HERE]

**Parent-gender differences**

Classification accuracy differed substantially between mothers and fathers across all models (Table 2). For the large reasoning model, accuracy was higher for mothers (93%) than fathers (85%). This eight percentage point gap persisted across model sizes: the medium model achieved 86% accuracy for mothers versus 82% for fathers, and the small model achieved 79% for mothers versus 81% for fathers. The RAG-based approach showed an even larger gap (85% for mothers, 75% for fathers).

Cohen's kappa values reflected this pattern (Table 4). For the large reasoning model, agreement with the benchmark was almost perfect for mothers ($\kappa = 0.85$) but only substantial for fathers ($\kappa = 0.66$). Error analysis revealed that the model failed to identify 16% (n = 5) of actual cases of paternal non-cooperation compared to 8% (n = 3) of maternal cases (Table 3). The model also produced more false positives for fathers (16.1%) than mothers (8.1%).

Notably, expert human reviewers exhibited similar difficulties. Prior to consensus resolution, inter-rater agreement between the two EHRs was substantially lower for fathers ($\kappa = 0.65$) than for mothers ($\kappa = 0.71$). The model's agreement with the benchmark for fathers ($\kappa = 0.66$) closely matched the agreement observed between the two human reviewers. These parallel patterns suggest that reduced classification accuracy for fathers reflects characteristics of the source documentation rather than model-specific limitations.

**Sensitivity analysis**

Sensitivity analyses compared model performance under the original three-category classification scheme versus the binary aggregation. Under the three-category scheme, overall accuracy for the large reasoning model was 78% compared to 89% under binary classification. The improvement reflected reduced ambiguity in distinguishing "cooperation present" from "no evidence" categories. Cohen's kappa showed a similar pattern (three-category: $\kappa = 0.65$; binary: $\kappa = 0.76$). The relative performance ranking across models remained unchanged under both classification schemes, indicating that the binary aggregation did not systematically advantage particular model configurations.

**Full corpus classification**

Table 5 presents classification results for the full corpus of 29,770 casework reports from 12,607 CPS cases using the large reasoning model. At the report level, lack of cooperation was identified in 17.5% of reports for mothers (n = 5,261) and 18.3% of reports for fathers (n = 5,483). At the case level, 17.6% of cases (n = 2,153) had at least one report documenting maternal non-cooperation, and 18.8% of cases (n = 2,366) had at least one report documenting paternal non-cooperation. Overall, 31.0% of cases (n = 3,900) had documented lack of cooperation by at least one parent at some point during CPS involvement.

[INSERT TABLE 5 ABOUT HERE]

Processing the full corpus required approximately 375 hours of computation time on high performance server-class hardware with 192GB of graphical memory, with each report requiring an average of 45 seconds for reasoning-based classification. This processing time, while substantial, enabled extraction of structured cooperation data from a corpus that would require an estimated 5,000 hours (approximately 120 weeks of full-time work) for manual review at ten minutes per report.

## Discussion

This study demonstrates that reasoning-enabled language models can accurately classify complex case factors in child protection documentation, including parental cooperation—an assessment domain characterized by ambiguous, temporally variable, and often contradictory information. Across all evaluated models, the large reasoning model achieved the strongest performance, improving accuracy from 80% (RAG-based approach) to 89% and approaching the inter-rater reliability of expert human reviewers. Accuracy scaled with model size, supporting the conclusion that greater reasoning capacity enhances performance for conceptually demanding tasks.

These results extend prior research showing that even small reasoning models can match or outperform larger instruction-tuned models on well-defined classification tasks such as substance-related problems or domestic violence (Qi et al., 2026). However, parental cooperation is conceptually distinct. It encompasses attitudinal, relational, and behavioral dimensions and unfolds over extended periods of child protection service involvement. Our findings indicate that larger reasoning models are better equipped to synthesize this complexity, particularly when integrating positive and negative evidence through structured chain-of-thought processing. The redesign of the prompting strategy, emphasizing explicit extraction of contradictory indicators and temporal trajectories and a comprehensive assessment of the case reports, substantially improved classification validity compared to earlier RAG-based methods (Stoll et al., 2025).

**Limitations**

Despite these advances, two methodological limitations warrant consideration. First, accuracy was consistently lower for fathers than mothers. The large reasoning model missed 16% of documented instances of paternal non-cooperation, compared to 8% for maternal cases. Agreement with expert human reviewers showed the same pattern and inter-rater reliability between human reviewers was likewise lower for fathers ($\kappa = 0.65$) than for mothers ($\kappa = 0.71$). These parallels suggest that the reduced accuracy is driven less by model behavior than by characteristics of the underlying documentation.

Second, the computational demands of the largest reasoning model remain substantial. Deploying a 255B-parameter reasoning model locally required specialized hardware (a minimum of 192GB of graphical memory) and generated an average inference time of 45 seconds per report. In contrast, 4–32 billion parameter reasoning models required only a fraction of the resources and achieved inference times of 4–8 seconds. Although these smaller models performed moderately well, the performance gap indicates that advanced reasoning models currently remain resource-intensive for large-scale operational use.

**Systemic bias in parental cooperation assessments**

Differences in documentation between mothers and fathers likely contribute to the performance gap. Across the full corpus, lack of cooperation was documented in 18.8% of cases for fathers and 17.6% for mothers. These figures should not be interpreted as true prevalence rates; automated classification can only assess what is written in the reports. Importantly, case documentation itself is shaped by biases that arise throughout the reporting process—such as uneven access to information, selective attention, memory distortions, and strategic choices about what to record. Because models can only analyze the available narrative text, these upstream biases directly influence how cooperation is represented for each parent and likely account for part of the observed difference between mothers and fathers. Prior research shows that child protection systems tend to be mother-centric, with mothers positioned as the primary point of engagement and fathers as peripheral figures (Dominelli et al., 2011; Philip et al., 2019; J. Scourfield et al., 2024; Strega et al., 2008). Documentation on fathers is not only less extensive but also framed in more dichotomous terms, limiting opportunities for nuanced assessment.

Our results reflect this pattern. Both human reviewers and reasoning models achieved lower accuracy when evaluating fathers, and both were more likely to miss indications of non-cooperation. This suggests that parental cooperation, as a documented case factor, cannot be measured equivalently across caregiver roles. The underlying concept appears implicitly gendered—rooted in different expectations for mothers and fathers—and these differences are embedded in professional assessments long before automated classification occurs. Thus, comparisons between maternal and paternal cooperation should be interpreted cautiously, as the construct itself may not be consistently operationalized in practice.

Automated classification does not eliminate such biases and may reinforce them if underlying documentation or model pre-training data contain systematic distortions (Garrido-Muñoz et al., 2021). We mitigated interpretive bias by defining cooperation constructs explicitly in the prompts and validating the reasoning process with human reviewers. Nevertheless, pre-training biases and asymmetries in source documentation remain unavoidable constraints. Future work should examine how documentation practices influence automated assessments and how model reasoning can be calibrated to identify and flag potential gaps or inconsistencies in narrative records.

## Conclusion

This study demonstrates that reasoning-enabled language models can accurately extract structured information about parental cooperation from CPS case reports, achieving reliability comparable to expert human reviewers even for conceptually complex and evolving case factors. By integrating chain-of-thought prompting with models designed for explicit reasoning, we developed a transparent and verifiable workflow that improves upon earlier RAG-based approaches and enables large-scale population-level analysis that would be infeasible through manual expert human assessment.

The findings underscore two broader implications. First, reasoning models provide a practical and methodologically robust approach for analyzing nuanced patterns in child welfare documentation, offering a viable pathway for generating structured data on factors that are not captured in administrative systems. Second, observed differences in model accuracy for mothers and fathers highlight that automated assessments inevitably mirror the biases embedded in professional practice and documentation. As such, reasoning-based classification not only offers analytic capability but also serves as a tool for revealing systemic blind spots that warrant further study.

Overall, the integration of advanced reasoning models into child welfare research holds promise for supporting more comprehensive, transparent, and evidence-informed assessments. Continued methodological refinement, combined with attention to gendered and structural biases in documentation, will be essential for ensuring that these tools contribute meaningfully to understanding and improving CPS practice.

# References

Ben-David, V. (2016). Parental Cooperation with Social Services and Termination of Parental Rights in Israeli Court Cases of Child Maltreatment. *Journal of Child and Family Studies*, *25*(8), 2498–2507. https://doi.org/10.1007/s10826-016-0422-9

Besta, M., Barth, J., Schreiber, E., Kubicek, A., Catarino, A., Gerstenberger, R., Nyczyk, P., Iff, P., Li, Y., Houliston, S., Sternal, T., Copik, M., Kwaśniewski, G., Müller, J., Flis, Ł., Eberhard, H., Chen, Z., Niewiadomski, H., & Hoefler, T. (2025). *Reasoning Language Models: A Blueprint* (Version 4). arXiv. https://doi.org/10.48550/ARXIV.2501.11223

Charest-Belzile, D., Drapeau, S., & Ivers, H. (2020). Parental engagement in child protection services: A multidimensional, longitudinal and interactive framework. *Children and Youth Services Review*, *116*, 105162. https://doi.org/10.1016/j.childyouth.2020.105162

Dettmers, T., Pagnoni, A., Holtzman, A., & Zettlemoyer, L. (2023). *QLoRA: Efficient Finetuning of Quantized LLMs*. arXiv. https://doi.org/10.48550/arXiv.2305.14314

Dominelli, L., Strega, S., Walmsley, C., Callahan, M., & Brown, L. (2011). 'Here's my Story': Fathers of 'Looked After' Children Recount their Experiences in the Canadian Child Welfare System. *British Journal of Social Work*, *41*(2), 351–367. https://doi.org/10.1093/bjsw/bcq099

Forrester, D., Westlake, D., & Glynn, G. (2012). Parental resistance and social worker skills: Towards a theory of motivational social work. *Child & Family Social Work*, *17*(2), 118–129. https://doi.org/10.1111/j.1365-2206.2012.00837.x

Garrido-Muñoz, I., Montejo-Ráez, A., Martínez-Santiago, F., & Ureña-López, L. A. (2021). A Survey on Bias in Deep NLP. *Applied Sciences*, *11*(7), 3184. https://doi.org/10.3390/app11073184

Gautschi, J. (with Pädagogischen Hochschule Freiburg). (2021). *Urteile und Entscheidungen unter Unsicherheit in Kindeswohlabklärungen. Einflussfaktoren auf Fallbeurteilungen in* *einer multifaktoriellen, experimentellen Vignettenstudie*. https://nbn-resolving.org/urn:nbn:de:bsz:frei129-opus4-8835

Jiang, A. Q., Sablayrolles, A., Mensch, A., Bamford, C., Chaplot, D. S., Casas, D. de las, Bressand, F., Lengyel, G., Lample, G., Saulnier, L., Lavaud, L. R., Lachaux, M.-A., Stock, P., Le Scao, T., Lavril, T., Wang, T., Lacroix, T., & Sayed, W. E. (2023). *Mistral 7B*. arXiv. https://doi.org/10.48550/arXiv.2310.06825

Jud, A., & Gartenhauser, R. (2015). The impact of socio-economic status and caregiver cooperation on school professionals' reports to child protection services in Switzerland. *European Journal of Social Work*, *18*(3), 340–353. https://doi.org/10.1080/13691457.2014.933093

Jud, A., Perrig-Chiello, P., & Voll, P. (2011). Less effort in worsening child protection cases? The time-course of intensity of services. *Children and Youth Services Review*, *33*(10), 2027–2033. https://doi.org/10.1016/j.childyouth.2011.05.032

Kwon, W., Li, Z., Zhuang, S., Sheng, Y., Zheng, L., Yu, C. H., Gonzalez, J. E., Zhang, H., & Stoica, I. (2023). *Efficient Memory Management for Large Language Model Serving with PagedAttention* (Version 1). arXiv. https://doi.org/10.48550/ARXIV.2309.06180

Landis, J. R., & Koch, G. G. (1977). The Measurement of Observer Agreement for Categorical Data. *Biometrics*, *33*(1), 159. https://doi.org/10.2307/2529310

Lätsch, D. C., Tausendfreund, T., & Brink, I. O. (2022). *Familiäre Ressourcen in der Krise? : Eine Studie zur Kinder- und Jugendhilfe des Kantons Zürich in Zeiten der Corona-Pandemie*. ZHAW Zürcher Hochschule für Angewandte Wissenschaften. https://doi.org/10.21256/zhaw-2429

Mason, C., Taggart, D., & Broadhurst, K. (2020). Parental Non-Engagement within Child Protection Services—How Can Understandings of Complex Trauma and Epistemic Trust Help? *Societies*, *10*(4), 93. https://doi.org/10.3390/soc10040093

Perron, B. E., Luan, H., Victor, B. G., Hiltz-Perron, O., & Ryan, J. (2024). Moving Beyond ChatGPT: Local Large Language Models (LLMs) and the Secure Analysis of Confidential Unstructured Text Data in Social Work Research. *Research on Social Work Practice*. https://doi.org/10.1177/10497315241280686

Philip, G., Clifton, J., & Brandon, M. (2019). The Trouble With Fathers: The Impact of Time and Gendered-Thinking on Working Relationships Between Fathers and Social Workers in Child Protection Practice in England. *Journal of Family Issues*, *40*(16), 2288–2309. https://doi.org/10.1177/0192513X18792682

Platt, D. (2012). Understanding parental engagement with child welfare services: An integrated model. *Child & Family Social Work*, *17*(2), 138–148. https://doi.org/10.1111/j.1365-2206.2012.00828.x

Qi, Z., Perron, B. E., Victor, B. G., Stoll, D., & Ryan, J. P. (2026). Small Models Achieve Large Language Model Performance: Evaluating Reasoning-Enabled AI for Secure Child Welfare Research. *Journal of Evidence-Based Social Work*, 1–22. https://doi.org/10.1080/26408066.2026.2616711

Raschka, S. (2025, December 30). The State Of LLMs 2025: Progress, Problems, and Predictions. *Ahead of AI*. https://magazine.sebastianraschka.com/p/state-of-llms-2025?utm_source=post-email-title&publication_id=1174659&post_id=182789318&utm_campaign=email-post-title&isFreemail=true&r=2cspju&triedRedirect=true&utm_medium=email

Scourfield, J. B. (2001). Constructing Men in Child Protection Work. *Men and Masculinities*, *4*(1), 70–89. https://doi.org/10.1177/1097184X01004001004

Scourfield, J., Davies, J., Jones, K., & Maxwell, N. (2024). Improving Children's Services Engagement of Fathers in Child Protection: Logic Model for an Organisational Development and Staff Training Intervention. *International Journal on Child

*Maltreatment: Research, Policy and Practice*, *7*(4), 607–614.

https://doi.org/10.1007/s42448-024-00206-y

Stoll, D., Jud, A., Wehrli, S., Lätsch, D., Steinmann, S., Wallimann, M. S., & Quehenberger, J. (2025). Case reports unlocked: Leveraging retrieval-augmented generation with large language models to advance research on psychological child maltreatment. *Child Abuse & Neglect*, *169*, 107653. https://doi.org/10.1016/j.chiabu.2025.107653

Strega, S., Fleet, C., Brown, L., Dominelli, L., Callahan, M., & Walmsley, C. (2008). Connecting father absence and mother blame in child welfare policies and practice. *Children and Youth Services Review*, *30*(7), 705–716. https://doi.org/10.1016/j.childyouth.2007.11.012

Swiss Civil Code (Zivilgesetzbuch, ZGB), Pub. L. No. Art. 307, SR 210 (1907). https://www.fedlex.admin.ch/eli/cc/24/233_245_233/en

Wei, J., Wang, X., Schuurmans, D., Bosma, M., Ichter, B., Xia, F., Chi, E., Le, Q., & Zhou, D. (2022). *Chain-of-Thought Prompting Elicits Reasoning in Large Language Models* (Version 6). arXiv. https://doi.org/10.48550/ARXIV.2201.11903

Witte, S. (2020). Case file analyses in child protection research: Review of methodological challenges and development of a framework. *Children and Youth Services Review*, *108*, 104551. https://doi.org/10.1016/j.childyouth.2019.104551

Yang, A., Li, A., Yang, B., Zhang, B., Hui, B., Zheng, B., Yu, B., Gao, C., Huang, C., Lv, C., Zheng, C., Liu, D., Zhou, F., Huang, F., Hu, F., Ge, H., Wei, H., Lin, H., Tang, J., … Qiu, Z. (2025). *Qwen3 Technical Report* (Version 1). arXiv. https://doi.org/10.48550/ARXIV.2505.09388

PARENTAL COOPERATION – RLMSTable 1 Data source meta information

| Data aspect | Value/Description |
| --- | --- |
| Number of public Child and Youth Welfare Agencies | 14 |
| Catchment area population (children and adolescents) | 220,000 |
| Number of counties served | 11 |
| Median duration of CPS assistance | 3.3 years |
| Upper-end quintile average duration CPS assistance | 10 years |
| Total number of CPS cases | 12,607 |
| Total number of casework reports (2008-2022) | 29,770 |
| Average number of reports per year | 2,200 |
| Average number of active cases per year | 4,600 |
| Average length of reports | 1,300 words (5 pages) |
| Standard deviation of report length | 730 words |

Table 2

Evaluation metrics for classification of parental cooperation compared to consensus dataset according to the applied model.

| Caregiver | Model | Metric | | | |
|---|---|---|---|---|---|
| | | Accuracy | Precision | Recall | F1 |
| Both Parents (N = 200) | Reasoning large (255B) | 0.89 | 0.89 | 0.89 | 0.89 |
| | Reasoning medium (32B) | 0.84 | 0.84 | 0.84 | 0.84 |
| | Reasoning small (4B) | 0.8 | 0.8 | 0.8 | 0.8 |
| | RAG Instruct large (123B) | 0.8 | 0.85 | 0.8 | 0.81 |
| Mother (N = 100) | Reasoning large (255B) | 0.93 | 0.93 | 0.93 | 0.93 |
| | Reasoning medium (32B) | 0.86 | 0.86 | 0.86 | 0.86 |
| | Reasoning small (4B) | 0.79 | 0.79 | 0.79 | 0.79 |
| | RAG Instruct large (123B) | 0.85 | 0.88 | 0.85 | 0.85 |
| Father (N = 100) | Reasoning large (255B) | 0.85 | 0.86 | 0.85 | 0.85 |
| | Reasoning medium (32B) | 0.82 | 0.83 | 0.82 | 0.82 |
| | Reasoning small (4B) | 0.81 | 0.81 | 0.81 | 0.81 |
| | RAG Instruct large (123B) | 0.75 | 0.82 | 0.75 | 0.76 |

Table 3

Confusion matrices for classification of parental cooperation compared to consensus dataset for the large reasoning model. Bold values indicate instances of false predictions, representing cases where the model's output did not match the actual classification.

| Caregiver | | Predicted False N (%) | Predicted True N (%) |
|---|---|---|---|
| Both Parents (N = 200) | Actual False | 60 (88.23%) | **8 (11.76%)** |
| | Actual True | **14 (10.60%)** | 118 (89.39%) |
| Mother (N = 100) | Actual False | 34 (91.89%) | **3 (8.10%)** |
| | Actual True | **4 (6.34%)** | 59 (93.65%) |
| Father (N = 100) | Actual False | 26 (83.87%) | **5 (16.12%)** |
| | Actual True | **10 (14.49%)** | 59 (85.50%) |

Table 4

Percentage agreement and Cohen's Kappa metric for inter-rater agreement for classification of parental cooperation between the model, expert human reviewers (EHRs) and the consensus dataset as a benchmark for the large reasoning model.

Values between k = .41 and k = .60 indicate moderate agreement, between k = .61 and k = .80 substantial agreement, and larger than k = .81 indicate complete agreement (Landis & Koch, 1977).

| Caregiver | | Percent Agreement | | | Cohen's Kappa | | |
|---|---|---|---|---|---|---|---|
| | | Model | EHR 1 | EHR 2 | Model | EHR 1 | EHR 2 |
| Both Parents (N = 200) | Model | | | | | | |
| | EHR 1 | 0.86 | | | 0.71 | | |
| | EHR 2 | 0.83 | 0.86 | | 0.62 | 0.68 | |
| | Benchmark | 0.89 | 0.94 | 0.91 | 0.76 | 0.88 | 0.80 |
| Mother (N = 100) | Model | | | | | | |
| | EHR 1 | 0.92 | | | 0.83 | | |
| | EHR 2 | 0.87 | 0.87 | | 0.71 | 0.71 | |
| | Benchmark | 0.93 | 0.97 | 0.90 | 0.85 | 0.94 | 0.77 |
| Father (N = 100) | Model | | | | | | |
| | EHR 1 | 0.81 | | | 0.59 | | |
| | EHR 2 | 0.80 | 0.85 | | 0.54 | 0.65 | |
| | Benchmark | 0.85 | 0.92 | 0.93 | 0.66 | 0.82 | 0.83 |

PARENTAL COOPERATION – RLMSTable 5

Results for parental lack of cooperation: Classification of casework reports and labelled CPS cases. Results are presented in absolute numbers and as percentage of all case reports and CPS cases, respectively.

| Level of Analysis | CPS Casework Reports | | Case | |
|---|---|---|---|---|
| Caregiver | *n* | % | *n* | % |
| Mother, lack of cooperation | 5,261 | 17.5 | 2,153 | 17.6 |
| Father, lack of cooperation | 5,483 | 18.3 | 2,366 | 18.8 |
| One of the parents, lack of cooperation | 9,367 | 31.3 | 3,900 | 31.0 |
| Total | 29,770 | 100.0 | 12,607 | 100.0 |

27